\documentclass[pre,onecolumn,superscriptaddress,aps,notitlepage,10pt]{revtex4-1}

\usepackage{mathtools}
\usepackage{newpxtext,newpxmath}
\usepackage[utf8]{inputenc}
\usepackage{hyperref}
\usepackage{graphicx}
\usepackage{lipsum}
\usepackage[normalem]{ulem}
\usepackage[margin=1.25in]{geometry}
\usepackage{subcaption}
\usepackage[dvipsnames]{xcolor}
\usepackage{quantikz}
\usetikzlibrary{shapes}
\usepackage{algorithm}
\usepackage{color}
\usepackage{algpseudocode}
\usepackage{comment}
\usepackage{enumitem}
\usepackage{url}
\usepackage{float}
\usepackage[textwidth=2.5cm]{todonotes}
\usepackage{adjustbox}
\usepackage{multirow}

\setuptodonotes{fancyline, color=blue!30}
\usepackage{outlines}

\begin{abstract}   
    Over the past decade, research in quantum computing has tended to fall into one of two camps: near-term intermediate scale quantum (NISQ) and fault-tolerant quantum computing (FTQC).
    Yet, a growing body of work has been investigating how to use quantum computers in transition between these two eras.
    This envisions operating with tens of thousands to millions of physical qubits, able to support fault-tolerant protocols, though operating close to the fault-tolerant threshold.
    Two challenges emerge from this picture: how to model the performance of devices that are continually improving and how to design algorithms to make the most use of these devices?
    In this work we develop a model for the performance of early fault-tolerant quantum computing (EFTQC) architectures and use this model to elucidate the regimes in which 
    algorithms suited to such architectures are
    advantageous.
    As a  concrete example, we show that, for the canonical task of phase estimation, in a regime of moderate scalability and using just over one million physical qubits, the ``reach'' of the quantum computer can be extended (compared to the standard approach) from 90-qubit instances to over 130-qubit instances
    using a simple early fault-tolerant quantum algorithm, which reduces the number of operations per circuit by a factor of 100
    and increases the number of circuit repetitions by a factor of 10,000.
    This clarifies the role that such algorithms might play in the era of limited-scalability quantum computing.
   
\end{abstract}

\begin{document}

\title{Early Fault-Tolerant Quantum Computing}

\author{Amara Katabarwa}
\affiliation{Zapata AI Inc., Boston, MA 02110 USA}

\author{Katerina Gratsea}
\affiliation{Zapata AI Inc., Boston, MA 02110 USA}
\affiliation{ICFO - Institut de Ci\`{e}ncies Fot\`{o}niques, The Barcelona Institute of Science and Technology, Av. Carl Friedrich Gauss 3, 08860 Castelldefels (Barcelona), Spain}

\author{Athena Caesura}
\affiliation{Zapata AI Inc., Boston, MA 02110 USA}

\author{Peter D. Johnson}
\affiliation{Zapata AI Inc., Boston, MA 02110 USA}

\maketitle

\section{Introduction}

Quantum computers were first proposed to efficiently simulate quantum systems \cite{Feynman1999SimulatingPW}. It then it took about a decade before it was discovered that quantum phenomena, such as superposition and entanglement, could be leveraged to provide an exponential advantage in performing tasks unrelated to quantum mechanics \cite{Deutsch1992}. Although of no practical use, the Deutsch–Jozsa algorithm sparked interest in using a quantum computer to perform other tasks beyond simulating quantum systems \cite{Grover1997, Bernstein1997}, the most famous case being Shor's algorithm  \cite{Shor1997}. 
Around the same time 
the ground-breaking discovery of quantum error correcting codes (QECC) \cite{Shor1995, Steane1997, Laflamme1996, Knill1996,Calderbank1997} set the stage for practical quantum computing.
This showed that errors due to faulty hardware could not only be identified but also corrected. Two pieces of the puzzle were left, namely: 
\begin{enumerate}
    \item Could quantum computation be done in a fault tolerant manner i.e. could error-corrected qubits perform better than physical qubits?
    \item Can one rigorously prove the existence of a threshold\footnote{A further requirement is that realistic assumptions be made on the noise model.} below which error can be reduced exponentially in the time and memory overhead cost?
\end{enumerate}
The first piece of the puzzle was tackled by Peter Shor \cite{Shor1996} and later, building on his work, threshold theorems were proved assuming various kinds of error models \cite{Aharonov2006, Aharonov2008, Kitaev2003}. For a specific quantum error correcting code and a noise model it is then left to prove and find error thresholds, with early works being \cite{aliferis2006simple, Aliferis2008, aliferis2008accuracy, Aliferis2006}; this continues to be an active area of research \cite{Fowler2012,Kovalev2013,Breuckmann2021, Cohen2022}.

Meanwhile on the hardware side, astonishing progress has been made 
across various
modalities (e.g. superconducting, ion trap, photonic, etc.) in terms of extending qubit coherence times and improving entangling operations \cite{Wintersperger2023, Shi2022, Debnath2016, Hanneke2009, Monroe1995, Bao2018, Chen2014}. Driven by
such advances,
a watershed moment occurred in 2016 when IBM put the first quantum computer on the cloud giving the public access to quantum computers. This event spurred widespread interest in finding near-term quantum algorithms that did not need the full machinery of fault tolerance. These algorithms first formulate the problem as a solution to the ground state of some Hamiltonian store a trial ansatz on the quantum processing unit (QPU) and use a classical optimizer to find the optimal parameters. The variational principle guarantees that the optimized parameters will produce a state whose energy upper bounds that of the target Hamiltonian. These so called hybrid \textit{quantum/classical} algorithms  allow one to use short depth quantum circuits and reduce the need for high quality quantum coherence. They have found application in areas of quantum chemistry  \cite{Peruzzo2014}, machine learning \cite{Liu2018, Dallaire-Demers2018a}  and optimization \cite{Farhi2014b}. 

Despite this progress, there is still need to reduce errors and the area of quantum error mitigation arose as attempts were made to meet the needs of these applications~\cite{Temme2017a, Czarnik2021, Giurgica-Tiron2020a, Huggins2021a, Berg2023}. This way of using a QPU is what is characteristic of the so called \textit{NISQ era} \cite{preskill2018quantum}. Although there is no strict definition of what constitutes a NISQ device it can generally be assumed that NISQ devices are too large to be simulated classically, but also too small to implement quantum error correction. IBM's work \cite{Kim2023} in some sense is the true dawn of the NISQ era i.e a quantum device where error mitigation is important and classical simulation is hard. The flurry of work \cite{Kechedzhi2023, Tindall2023, Tomislav2023} immediately arose pushing classical methods of simulation and claiming to reproduce IBM's results. This is a new phase in which NISQ devices will be put to the test by state of the art classical simulators and vice-versa. This back and forth will not last long as Hilbert space for quantum systems grows exponentially and the NISQ device will be the only viable simulation approach.

But an important question remains and in a very obvious sense the elephant in the room is, ``are NISQ devices and NISQ algorithms up for the task of realizing quantum advantage at utility scale?'' Work has been done in quantum chemistry where the problem can be precisely asked, for example, in finding the ground state of large molecules. The best estimates so far for resource estimates suggest the variational quantum eigensolver (VQE) is not up to the task \cite{gonthier2022measurements}. Other work suggests a possible quantum advantage for the quantum adiabatic optimization ansatz (QAOA) \cite{Farhi2016, Shaydulin2023, Lykov2023} in optimization but it remains to be seen whether these claims can be confirmed in the presence of noise at scale. 

Given these roadblocks, should our attitude be to wait for fully fault tolerant devices? An area of research offers an intriguing possibility; we are offered a trade-off, we require fault tolerant quantum computing but the ability to run smaller quantum circuits at the cost of requiring more sampling for the quantum device. 
Such a trade-off has been the focus of a substantial amount of research in the past few years \cite{wang2019accelerated, parrish2019quantum, wang2021minimizing, giurgica2022lowa, klymko2022real, wang2023quantum, dong2022ground, wang2022state, ding2022even, kirby2023exact}
However, in a regime where we are able to arbitrarily scale the number of physical qubits while maintaining quality fault-tolerant protocols, such a trade-off would not be favorable; by increasing the circuit size using methods such as quantum amplitude amplification \cite{cleve1998quantum}, the additional overhead of efficient fault-tolerant protocols is negligible compared to the overall reduction in runtime.
Accordingly, such a trade-off would be better suited
to a setting
in which the efficiency of fault-tolerant protocols worsens with increasing system size.
If the ability to scale the number of physical qubits (i.e. the ``scalability'') is compromised by a worsening of the operations, then this diminishing returns will, in turn, limit size of problems that can be solved.
Such a regime of computation has been referred to as \textit{early fault tolerant quantum computing} (EFTQC) \cite{campbell2021early}, a natural successor to the NISQ era. 
A field of research has emerged recently where the proposed quantum algorithms enable this ``circuit-size vs sample-cost'' trade-off
\cite{wang2019accelerated, Zhang2021, wang2021minimizing, lin2022heisenberg, wan2022randomized, giurgica2022lowa,
dong2022ground, Ding2023, Wang2022, Wang2023}.
Two questions are then placed before us: 
\begin{enumerate}
    \item Will this regime of limited-scale quantum computers exist in a meaningful way?
    \item If so, will we be able to unlock intrinsic quantum value at scale  in this regime?
\end{enumerate}

\begin{figure}[ht]
    \centering
    \includegraphics[width=0.6\textwidth]{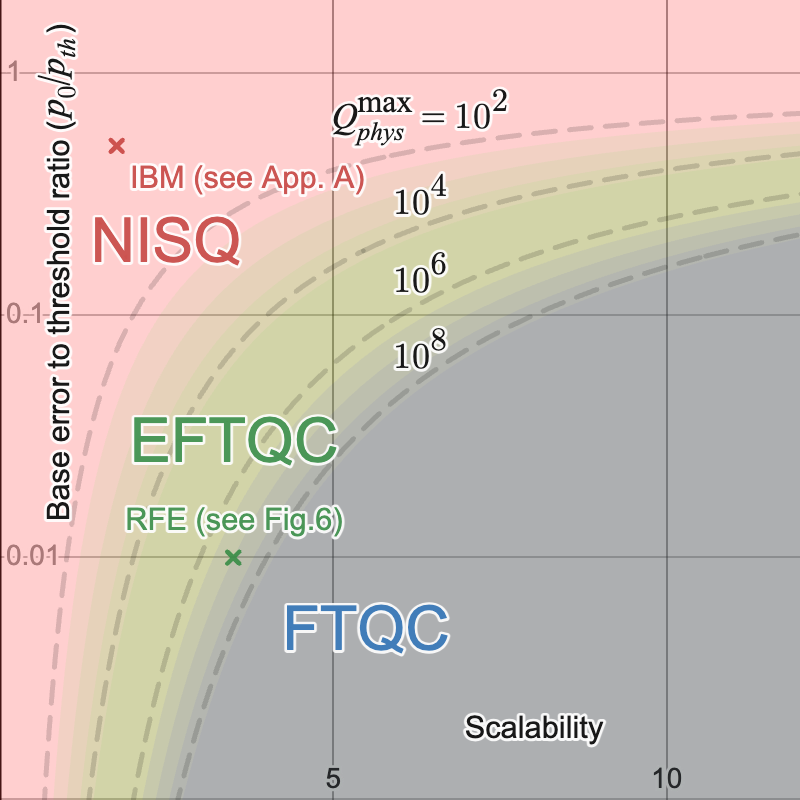}
    \caption{This figure roughly demarcates the regimes of NISQ, EFTQC, and FTQC according to the scalability model introduced in Section \ref{sec:model}. The vertical axis quantifies the base error rate (i.e. that achievable for a single-qubit), while the horizontal axis quantifies the ability of the architecture to maintain low error rates as it is scaled (i.e. its \emph{scalability}).
    Contours indicate the maximum physical qubit number that the architecture is warranted in scaling to as predicted by the scalability model of Equation \ref{eq:scalability}. The NISQ-to-EFTQC transition is characterized by having enough qubits to implement fault-tolerant non-Clifford operations (e.g. T factories), while the EFTQC-to-FTQC transition is characterized by the ability to accommodate very large problem instances (e.g. encoding 10,000 logical qubits using in $10^9$ physical qubits). The red x corresponds to data presented in Section \ref{app:ibm}, which estimates that a hardware vendor of today (IBM) has a scalability of 1.75 with $p_0=0.005$. An editable version of the plot can be accessed here: \href{https://www.desmos.com/calculator/9iphmmdjfp}{https://www.desmos.com/calculator/9iphmmdjfp}}
    \label{fig:regimes}
\end{figure}

The ultimate answers to these questions will depend on hard-to-predict factors including hardware, QECC and quantum algorithm advances, and improvements in competing classical hardware and algorithms.
Rather than predicting the timeline of these advances, we propose a quantitative framework to track their progress. 
Figure \ref{fig:regimes} depicts the landscape in which this framework assesses the ability of a given hardware vendor to supply useful physical qubits, transitioning from NISQ to EFTQC to FTQC.

To address the first question we propose a very simple model (see Equation \ref{eq:scalability}) to quantitatively discuss these regimes.
This simple model describes how the quality of elementary quantum operations degrade as system size is increased; that is, we model the physical gate error rate as a function of physical qubit number.  We dub this model the \textit{scalability} of a device.
For the second question, we quantify how recently-developed algorithms can extend the ``reach'' of quantum computers with limited \textit{scalability}. 
This is an important step towards understanding what value such methods can provide.
Two results that we will establish (see Equations \ref{eq:QPopt} and \ref{eq:QLmaxbound}) are that, according to the scalability model, the optimal number of physical qubits to use in the architecture is the following function of scalability parameter $s$
\begin{align}
Q_{\textup{phys}}^{\textup{opt}}=\frac{1}{e^2}\left(\frac{p_{\textup{th}}}{p_0}\right)^s
\end{align}
and the maximum problem size, measured in terms of the largest number of logical qubits, is predicted to be
\begin{align}
Q_L^{\textup{max}}\approx\frac{\left(\frac{p_{\textup{th}}}{p_\textup{phys}}\right)^s}{2e^2s^2\beta^2\ln\left(\left(\frac{A\alpha}{p_{C}}\right)^{\frac{1}{\beta}}\frac{\left(\frac{p_{\textup{th}}}{p_\textup{phys}}\right)^s}{8e^2s^2\beta^2}\right)^2}.
\end{align}
The various parameters are defined in Section \ref{subsec:qpe}.
We will ultimately explore so-called ``EFTQC algorithms'', which enable an increase in $Q_L^{\textup{max}}$.
We will explain how these expressions show 1) the importance of the scalability parameter $s$ in governing the capabilities of a quantum hardware vendor and 2) the role played by the ``fault-tolerance burden factor'' $\frac{A\alpha}{p_{C}}$, that combines gate count, error correction, and algorithm robustness parameters.
These elucidate multiple ways to improve a quantum computation towards solving utility-scale problem instances in the finite-scalability regime.

The manuscript is organized as follows.
In Section \ref{sec:model} we present the scalability model and apply this to an example resource estimation for the quantum phase estimation algorithm.
In Section \ref{sec:eftqc} we review progress in algorithms for early fault-tolerant quantum computers and then present an example of one such algorithm, showing how can improve the capabilities of a device with limited scalability.
Finally, in Section \ref{sec:discussion} we discuss the implications of our findings
and outline important future research directions.
 
\section{Modeling Early Fault-tolerant Quantum Computations}
\label{sec:model}
    
\subsection{Introduction to the scalability model}\label{subsec:scalability}

In this section we establish and discuss the precise sense in which a device can be an \textit{early fault tolerant device}. We first note the tension in the very phrase \textit{early fault tolerance}.
Fault-tolerance evokes the ability to ensure efficient suppression of error despite the use of faulty operations~\cite{Shor1996}. The string of results \cite{Dorit2008, Knill1996, aliferis2008accuracy, Aharonov2006} 
collectively known as the threshold theorems show that  \emph{in principle} this can be achieved. In fact thanks to these results we know~\cite{Aharonov2006} that under quite general assumptions such as allowing for long range correlations of noise and non-Markovianity, fault-tolerance is still possible. These foundational works would put the threshold error rate around  $10^{-5}$ to $10^{-6}$. However, more optimistic threshold predictions have been made using numerical investigations \cite{zalka1996threshold}.
For the surface code~\cite{dennis2002topological}, which is a leading contender for practical quantum computing~\cite{Google2022}, such simulations have led to the prediction of quite optimistic thresholds of $\sim 1\%$ \cite{wang2011surface}, which have also been argued for analytically \cite{fowler2012proof}. 
On the other hand,  numerical thresholds are based on particular assumptions of noise and error that cannot fully capture the complexity of quantum architectures at scale.
For example, an important assumption is that a single number can be used to capture the performance of operations and that this single number remains constant as larger code distances\footnote{In the surface code, the degree of resilience to error is controlled by the size of the two-dimensional grid of qubits used to encode each qubit. The minimal number of single-qubit errors needed to cause a logical error, or the code distance, is the diameter of the two-dimensional grid of qubits.} are used \cite{wang2011surface}.
Such thresholds have become the established targets for hardware developers~\cite{Xue2022, Blume-Kohout2017, Postler2022, Egan2021}.

The "early" in "early fault tolerance" on the other hand suggests some kind of limited ability to achieve fault tolerance i.e using polynomial amount of resources to achieve exponential error suppression\cite{Gottesman1997, Raussendorf2006}. This tension is what lies behind the motivation for this work.

A key insight towards resolving this tension is to realize what we might call the \textit{scalability requirement}:

\begin{quote}
    \textit{In order to reap the benefits of being below any threshold, an approach to building a quantum architecture must be able to maintain each operation below the threshold error rate as larger and larger architectures are built.}
\end{quote}

The failure to achieve the scalability requirement implies the existence of scale dependent errors. To motivate where these scale dependent errors might come from we consider the general setup used to prove the threshold theorem. It is assumed we have the following Hamiltonian as     
\begin{equation*}
    \mathcal{H} = \mathcal{H}_S + \mathcal{H}_B + \mathcal{H}_{SB}, 
\end{equation*}
where $\mathcal{H}_s$ is the hamiltonian governing the evolution of the system which for our discussion can be the evolution corresponding to implementing the quantum gate, $\mathcal{H}_B$ governs the evolution of some bath and $\mathcal{H}_{SB}$ entangles the bath with the qubits in the computation. 
The scale dependent errors arise from the engineering details involved in implementing $\mathcal{H}_S$ as larger and larger chips are developed. These engineering problems can't be completely inserted into $\mathcal{H}_{SB}$ and yet would ultimately impact how easily we could stay below threshold as we try to scale up. For fixed frequency qubits in superconducting architectures the issue of "frequency crowding" affects the quality that any single two-qubit gate can achieve \cite{Hutchings2017}. The number of frequencies that must be avoided when implementing the cross resonant gate, increases as the number of qubits increases in the chip; this makes targeting the required frequency harder and harder as you scale up. Another scale dependent engineering difficulty can arise from unwanted interactions between control lines going into the chip. The calibration of these pulses is partly a classical problem that gets more complicated and cumbersome as the chip gets larger. The issue of "cross mode coupling" at ion-traps will affect the fidelity of the gate \cite{Leung2018, Liang2023}, where the target has a specific motional mode but unwanted couplings destroy the quality of the gate. The problem has a classical component that scales with the number of qubits. In the above cases the physics of accurate addressability of a qubit or pairs of qubits is a problem that becomes harder with increasing number of qubits, and thus, affects the quality of the gate operation. 
Recent works have explored the consequences of scale dependent errors which would most likely arise from the limited resources to control qubits and design good quality operations \cite{Fellous-Asiani2021, Fellous2021}.

It is reasonable to believe that the assumption of scale-independent error rates may eventually become effectively true on account of modularity, as future quantum architectures will likely be made from repeated modular components.
And while the holy-grail of (effectively) scale-independent, sub-threshold error rates may someday be realized, quantum architectures will necessarily undergo a transition from today's scale-dependent error to the future of scale-independent error.
We will take this transition to be the defining characteristic of early fault-tolerant quantum computing.

Ultimately, this investigation is motivated by wanting to understand the prospects of using early fault-tolerant quantum computers to solve utility-scale problems.
We take such machines to be characterized by a non-negligible degree of scale-dependent error.
The standard approach to predicting the performance of fault-tolerant architectures for utility-scale problems is to assume that the error is scale independent~\cite{kim2022fault, goings2022reliably, beverland2022assessing}.
Therefore, our approach will be seen as 1) a generalization that incorporates both the scale-independent and dependent settings and 2) an attempt to bridge the observed scale-dependence of error in today's devices with the hoped-for scale-independence of error in future quantum architectures.
We expect that the degree of scale-dependence will inform the capabilities of the architecture being modeled.
Furthermore, scale-dependent error may warrant the development and use of quantum algorithms that are suited to this limitation.
These considerations motivate the main question pursued in the remainder of this manuscript:
\emph{how does the degree of scale-dependent error determine the capabilities of early fault-tolerant quantum computers?}
Next, we introduce a model to capture the degree of scale-dependent error.

We start by describing the particular setting in which we model scale-dependent error.
Our model will center around the concept of \emph{scalability}, the ability to maintain low error rates (e.g. sub-threshold) as larger architectures are requested.
Our setting and model are driven by the need to answer the question: for a \emph{series} of quantum computations of increasing size, how well will a hardware vendor be able to service the request to run the quantum computations.
Accordingly, we will not consider the capabilities of a single quantum device or a single quantum architecture, as the hardware vendor might have several architectures to service computations of various sizes. 
Furthermore, we will not consider the capabilities of the hardware vendor as they improve over time, as our hypothetical test is used to assess capability at one moment in time. 

\begin{figure}[ht]
    \centering
    \adjustbox{cfbox=black 0.25pt}{\includegraphics[width=70mm, height=65mm]{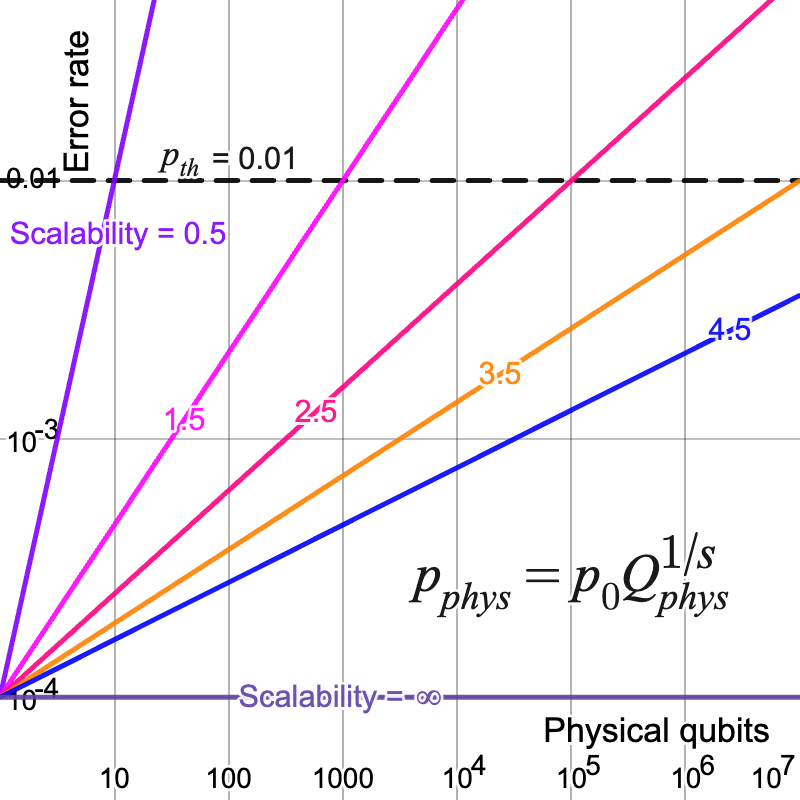}}
    \caption{The concept of scalability captures the ability of a quantum architecture to maintain low physical error rates as the number of physical qubits of the architecture is increased. This figure shows the scalability profiles of different quantum architectures given by the scalability model (Eq.~\ref{eq:scalability}) for different scalability values ($s=0.5, 1.5, 2.5, 3.5, 4.5, \infty$) and base error rate $p_0=10^{-4}$. A finite scalability implies that beyond a certain physical qubit size, the architecture cannot maintain physical error rates below the error threshold ($p_{th}$) of the fault-tolerant protocol.
    An editable version of the plot can be accessed here: \href{ https://www.desmos.com/calculator/jlmbygcqrp}{ https://www.desmos.com/calculator/jlmbygcqrp}}
    \label{fig:scalability_k}
\end{figure}
In order to make this quantitative, we can consider a \emph{scalability profile}: an empirically derived function that reports the worst-case error rate among the elementary operations of the device as a function of the requested number of physical qubits.
For the case of today's IBM devices, we present data on their scalability profile in Appendix \ref{app:ibm}.
In lieu of scalability profile data for future quantum vendors, we propose a simple parameterized model for this function

\begin{equation}
    \label{eq:scalability}
    p_{\textup{phys}}(Q_{\textup{phys}};\mathcal{V}) = p_0 Q_{\textup{phys}}^{1/s}, 
\end{equation}
where $Q_{\textup{phys}}$ is the number of physical qubits in the architecture and $\mathcal{V}$ labels the particular hardware vendor that is providing the qubits at any time. Parameters $p_0$ and $s$ capture the base error rate and the ``scalability'', respectively. 
It is helpful to view this model as a power-law fit of a scalability profile.
In Appendix~\ref{app:log_model} we investigate the more optimistic case of a logarithmic model.
The case of $s=\infty$ corresponds to scale-independent error, or infinite scalability, while any finite value of $s$ corresponds to the case of finite or limited scalability.

As we will show in the next section, in the context of fault-tolerant quantum computing, a finite scalability will result in a finite limit on the number of physical qubits being used before fault-tolerant protocols yield diminishing returns. We then explain how this limit on physical qubit number places a limit on the problem sizes that the architecture can accommodate.
Importantly, all of these considerations apply in the setting where fault-tolerant protocols are being used.
This differs from the setting assumed for NISQ quantum computing \cite{preskill2018quantum}, where physical qubits instead of logical qubits are used for computation.
Before moving to the next section, we provide some perspective on the transition from the NISQ regime to the EFTQC regime.
Specifically, in the rest of this subsection 
we estimate the minimal number
for $Q_{\textup{phys}}$ in an EFTQC computation assuming a simple surface code architecture.

The total number of physical qubits for a computation can be written as $Q_{\textup{phys}} = Q_{\textup{comp}} + Q_{\textup{MSD}}$ where $Q_{\textup{comp}}$ is the number of physical qubits used to compute (i.e. storing and routing the logical data) and $Q_{\textup{MSD}}$ are the physical qubits used for magic state distillation. To calculate the minimum number of qubits required for QEC, we will set $Q_{\textup{comp}} = 2 (d+1)^2$~\cite{goings2022reliably} corresponding to single surface code logical qubit and pick the smallest distillation widgets which give an improvement on the error rate. 

The most efficient distillation widgets known in the surface code are given in~\cite{litinski2019magic}. We have listed the smallest of these in Figure~\ref{fig:small_msd} (note that these do not give much of an improvement over the physical error rate). 
An important property of the magic state injection process is that it cannot have error rates which are less than the current logical level. Thus a magic state which is injected into a code with error rate $p_L$ can have at most error rate $p_L$. If we have a single logical qubit in the smallest non-trivial surface code (i.e. distance 3), then this minimum viable example of EFTQC will be at least 540 and 826 qubits for lower and high-quality operations,
respectively.

Note that the magic state distillation factory dominated the number of qubits. As a result, the FTQC community has put a lot of work into decreasing the size of factories~\cite{litinski2019magic}, improving injection protocols\cite{gavriel2022transversal},  or eliminating distillation entirely~\cite{akahoshi2023partially}. One would expect that the first EFTQC demonstrations will employ many of these techniques rather than the "pure FTQC" calculation presented above. In a more careful calculation to estimate a lower bound for the EFTQC range, one may want to take such techniques into account and calculate $Q_{\textup{phys}}$.
Refining this estimate to clarify and lower the NISQ-to-EFTQC transition is important future work.

\begin{figure}
    \centering
    \begin{tabular}{c|c|c|c|c|c|c}
         Quality of operations & Factory name & $p_{\textup{phys}}$ & $Q_{\textup{phys}}$ & $p_{\textup{out}}$ & $Q_{\textup{min, EFTQC}}$ & $p_L$\\\hline
         High & $\text{(15-to-1)}_{5,3,3}$ & $10^{-4}$ & 522 & $4.7 \times 10^{-6}$ & 554 & $10^{-5}$\\
         Lower &$\text{(15-to-1)}_{7,3,3}$ & $10^{-3}$ & 810 & $5.4 \times 10^{-4}$ & 842 & $10^{-3}$\\
    \end{tabular}
    \caption{A table containing the two smallest possible magic state distillation factories given by~\cite{litinski2019magic}. $p_{\textup{phys}}$ is the physical error rate,  $Q_{\textup{phys}}$ is the number of physical qubits required to create the factory, $p_{\textup{out}}$ is the probability that the output state magic state is incorrect, $Q_{\textup{min, EFTQC}}$ is a rough lower bound on the number of qubits in an EFTQC calculation, and $p_L$ is the logical failure rate in that lower bound calculation. Note that in the case of superconducting qubits, the lower bound EFTQC example does not decrease the logical error rate $p_L$. 
    }
    \label{fig:small_msd}
\end{figure}

\subsection{Example: quantum phase estimation compiled to the surface code}
\label{subsec:qpe}

\begin{figure}[ht]
    \centering
    \adjustbox{cfbox=black 0.25pt}{\includegraphics[width=0.6\textwidth]{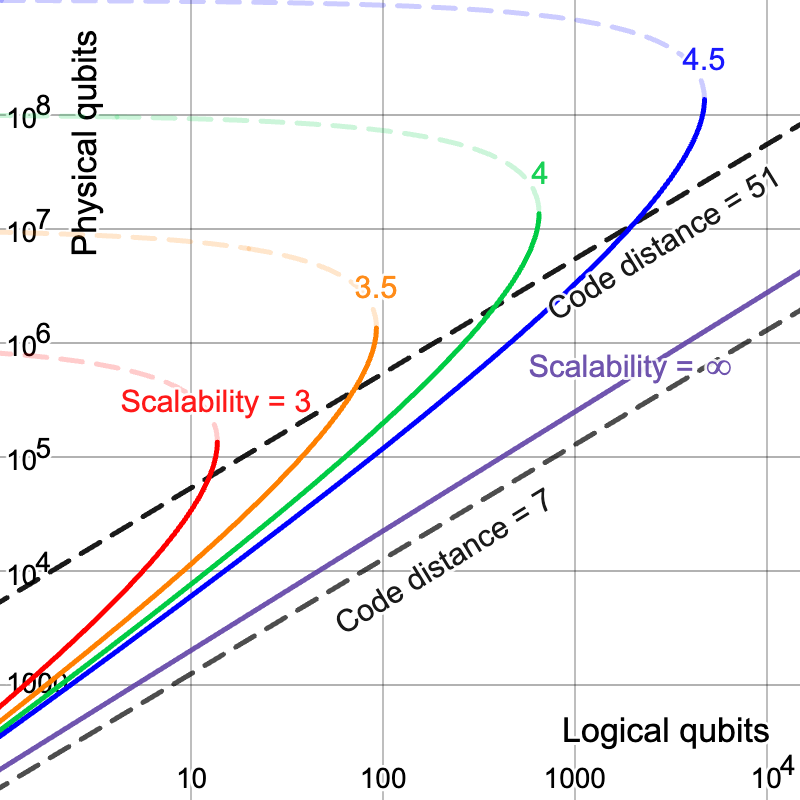}}
    \caption{The scalability model of Eq. \ref{eq:scalability} predicts that, for each finite value of scalability parameter $s$, there is a maximum problem instance size that can be accommodated by the architecture.
    Each curve is a contour in the $Q_{\textup{phys}}$-$Q_L$ plane of a solution to Eq. \ref{eq:success_condition} for a particular value of the scalability parameter $s$ (3, 3.5, 4, 4.5, $\infty$).
    The remaining parameters of Eq. \ref{eq:success_condition} are set to $p_{th}=10^{-2}$, $p_{0}=10^{-4}$, $\alpha=4.12\cdot 10^9$, $\beta=0.515$ following Table II in \cite{campbell2021early}.
    The transition from solid to faded dashed curves occurs when the physical qubit number reaches $Q_{\textup{phys}}^{\textup{opt}}=Q_{\textup{phys}}^{\textup{max}}/e^2$, beyond which increasing the code distance leads to diminishing returns. 
    The diagonal black dotted lines show the physical qubit count for two fixed code distances: 7 (small distance) and 51 (large distance). Note that code distance is discrete, which, if taken into account, would result in the contours jumping from one fixed-code-distance line to the next. 
    However, we have chosen to allow for the distance parameter to be continuous, for ease of viewing the trends of the contours.
    An editable version of the plot can be accessed here: \href{https://www.desmos.com/calculator/7mbziuf8gd}{https://www.desmos.com/calculator/7mbziuf8gd}}
\label{fig:scalability}
\end{figure}

In the preceding subsection, we introduced Eq. \ref{eq:scalability} as a model for how physical operation error rates might increase with system size. To understand the implications of this model, we work through the example of using the quantum phase estimation (QPE) algorithm \cite{cleve1998quantum} to solve the phase estimation task.
The task of phase estimation is to estimate the eigenphase of a unitary operator \(U\) with respect to an eigenstate \(\ket{\psi}\) assuming access to circuits that implement $c$-$U$ and prepare $\ket{\psi}$.
We review how to estimate the quantum resources required to perform this task under the scalability model and compare these to the ideal model case (i.e. $s\rightarrow \infty$).

A fault-tolerant resource estimation answers the question: how many physical qubits are needed per logical qubit to ensure that the logical error rates are sufficiently low to make the algorithm succeed (with some probability)?
To answer this, we must 1) determine what logical error rates the algorithm deems as ``sufficiently low'' and 2) establish the relationship between logical error rate and quantum resources.

For 1), the QPE algorithm will succeed with sufficiently high probability as long as the total circuit error rate is below some value $p_C$. We will set $p_C = 0.1$, noting that, in the literature, this tolerable circuit error rate varies from $0.1$ \cite{goings2022reliably} to $0.01$ \cite{kim2022fault}, but can be made lower using alternative algorithms \cite{kimmel2015robust, kshirsagar2022proving, li2023low}.
This tolerable circuit error rate, along with the number of operations per circuit, lets us bound the tolerable operation error rate. 
The quantum circuit will ultimately be compiled into a set of logical operations that are implemented using fault-tolerant protocols (e.g. initialization of $\ket{0}$, measurement in the computational basis, $H$ gate, CNOT gate, and $T$ gate).
We define $G_C$ to be the number of elementary logical operations (including idling\footnote{Especially in the case where the quantum computation is rate-limited by magic state distillation, the computational qubits would be required to idle without accruing error while waiting for T gates or Toffoli gates to be teleported into the computation.}) used by the circuit.
To ensure that the circuit error rate is less than $p_C$, it suffices\footnote{In the case where the tolerable circuit error rate approaches 1, $p_L \leq (1/G_C)\ln (1/(1-p_C))$ can be used as a tighter bound.} to ensure a logical error rate of $p_L \leq p_C/G_C$ (by the union bound).

For the quantum phase estimation algorithm, $G_C$ is determined by the target accuracy and the number of operations per $c$-$U$.
To yield an estimate of the phase angle to within \(\epsilon\) of the true value requires using a circuit with \(1/\epsilon\) applications of $c$-$U$~\cite{brassard2002quantum}. 
For our purposes we assume a model for $G_C$ by fitting data in Table II of \cite{campbell2021early} to the following power law, where the $\epsilon$ is set to be approximately half a percent of the total system energy,
\begin{align}
    G_C = \alpha Q_L^\beta,
\end{align} yielding $\alpha = 4.12\cdot 10^9$ and $\beta\approx 0.515$.
Thus, the algorithm success is ensured (with high probability) by
\begin{align}
\label{eq:alg_success}
    p_L \leq \frac{p_C}{\alpha Q_L^\beta}.
\end{align}
For simplicity, we'll assume that the number of logical qubits needed for magic state factories are accounted for in this model (see notes in the \href{https://www.desmos.com/calculator/7mbziuf8gd}{desmos plot} of Figure \ref{fig:scalability} for details of the assumptions and the relevant references) and we will assume that the physical qubit overhead is captured by the code distance used for the data qubits (though the factories typically have multiple layers of concatenation with differing code distance).

For 2) we will assume a model of error suppression based on simulations of the surface code in \cite{fowler2018low}.
This model is
\begin{align}
    p_L = A (p_{\textup{phys}}/p_{\textup{th}})^{(d+1)/2},
\end{align}
where \cite{fowler2018low, goings2022reliably} estimates $A=0.1$ and $p_{\textup{th}}=0.01$.
The number of physical qubits used to encode one logical qubit in the surface code is $2(d+1)^2$, leading to
\begin{align}
Q_{\textup{phys}}=2(d+1)^2Q_L.
\end{align}
In the case that $p_{\textup{phys}}$ is independent of the number of physical qubits, $p_L$ can be made arbitrarily small, with cost (depending on code distance $d$) scaling as $d\sim \log(1/p_L)$. 
However, if we replace $p_{\textup{phys}}$ with the $Q_{\textup{phys}}$-dependent function $p_{\textup{phys}}(Q_{\textup{phys}})$ of Eq. \ref{eq:scalability} (i.e. the scalability model), the logical error rates cannot be made arbitrarily small.
The smallest error rate is achieved when $p_{\textup{phys}}=p_{\textup{th}}$, which occurs when $Q_{\textup{phys}} = (p_{th}/p_0)^s$;
including more qubits (i.e. increasing the code distance) will lead to a decrease in the logical error rate.
This number of physical qubits is therefore the maximal number of physical qubits that should be used under the scalability model: 
\begin{align}
\label{eq:QPmax}
    Q_{\textup{phys}}^{\textup{max}} = (p_{th}/p_0)^s.
\end{align}
So, for example, when $p_{th}=0.01$ and $p_0=0.001$ (as is sometimes assumed for superconducting qubit resource estimates with the surface code \cite{goings2022reliably}) we have $Q_{\textup{phys}}^{\textup{max}}=10^s$.
A more optimistic setting of $p_0=0.0001$ leads to $Q_{\textup{phys}}^{\textup{max}}=10^{2s}$.
Figure \ref{fig:regimes} depicts contours of $Q_{\textup{phys}}^{\textup{max}}$ in the plane of $p_0/p_{th}$ vs $s$.

The above concepts can be summarized as follows:
\[\begin{array}{ll}
\textbf{Requirement} \\
p_C \geq G_Cp_L  & (\text{Algorithm Error Tolerance}) \\[2em]
\textbf{Cost} \\
Q_{\textup{phys}} = 2(d+1)^2 Q_L & (\text{Surface Code Overhead}) \\[2em]
\textbf{Models} \\
G_C = \alpha Q_L^\beta & (\text{QPE Circuit Gate Count}) \\
p_L = A \left(\frac{p_{\textup{phys}}}{p_{\textup{th}}}\right)^{\frac{d+1}{2}} & (\text{Surface Code Logical Error Rate}) \\
p_{\textup{phys}} = p_0 Q_{\textup{phys}}^{1/s} & (\text{Scalability Physical Error Rate Model}) \\[2em]
\end{array}
\]
Putting these together, we can determine the number of physical qubits required to ensure that QPE returns an $\epsilon$-accurate estimate (with high probability) as a function of the number of logical qubits $Q_L$ (roughly corresponding to problem size).
This relationship is expressed by $Q_{\textup{phys}}$-$Q_L$ pairs that ensure Eq. \ref{eq:alg_success} is satisfied (i.e. that logical error rates are low enough for the algorithm to succeed),
\begin{align}
\label{eq:success_condition}
\sqrt{8Q_L}\log\left(\frac{A\alpha}{p_C} Q_L^\beta\right)\leq \sqrt{Q_{\textup{phys}}}\log \left(\frac{p_{\textup{th}}}{p_{0}}Q_{\textup{phys}}^{-1/s}\right).
\end{align}
Before applying this result to the quantitative example that has been set up, we make a few general remarks that apply to any algorithm analyzed in this manner.

First, we consider the right-hand side of this inequality. This function will determine an optimal value for $Q_{\textup{phys}}$, which we label as $Q^{\textup{opt}}_{\textup{phys}}$ . Previously, we described a maximum value of $Q_{\textup{phys}}$ as set by the condition of $p_{\textup{phys}}$ being below threshold.
However, the maximum allowed value of $Q_L$ is now set by a function of $Q_{\textup{phys}}$; to increase this ceiling, we should maximize the right hand side function of $Q_{\textup{phys}}$. This function achieves its maximum of $\left(\frac{2e}{s}\right)^2\left(\frac{p_{\textup{th}}}{p_0}\right)^s$ at a value of 
\begin{align}
\label{eq:QPopt}
Q^{\textup{opt}}_{\textup{phys}}=\frac{1}{e^2}\left(\frac{p_{\textup{th}}}{p_0}\right)^s\leq Q^{\textup{max}}_{\textup{phys}}.
\end{align} 
This is considered the optimal number of physical qubits in that it enables the use of the largest number of logical qubits.
As an example, for $p_{\textup{th}}=0.01$, $p_{0}=0.0001$, and $s=3.5$, the optimal number of physical qubits is
$Q^{\textup{opt}}_{\textup{phys}}\approx 1.35 \times 10^6$.

These quantities of $Q^{\textup{max}}_{\textup{phys}}$ and $Q^{\textup{opt}}_{\textup{phys}}$ can help us to quantify the scalability parameters $p_0$ and $s$ that are relevant to the NISQ-to-EFTQC transition and the EFTQC-to-FTQC transitions.
At the end of the previous subsection we described how the NISQ-to-EFTQC might occur in the range of 100 to 10,000 physical qubits.
Considering Equation \ref{eq:QPmax} and \ref{eq:QPopt}, this determines the ($p_0$,$s$) pairs characteristic of this transition and shown as the red-to-green blend in Figure \ref{fig:regimes}.

We motivate the idea that the transition from EFTQC to FTQC is characterized by how the quantum computations are ``bottle-necked''.
In the case of fault-tolerant quantum computing, it is envisioned that the ability to run larger and larger quantum computations is possible as long as the computations are not practically limited by resources such as time and energy.
We propose that early fault-tolerant quantum computing be characterized by the regime in which the largest possible quantum computations are limited by the maximum number of physical qubits warranted in the architecture ($Q^{\textup{max}}_{\textup{phys}}$ or $Q^{\textup{opt}}_{\textup{phys}}$).
Viewing time as the limiting resource, if we assume that the quantum computation must finish within a month, then this limits the problem sizes that can be accommodated accordingly.
Using the quantum chemistry resource estimations of \cite{goings2022reliably} as a point of reference, problem instances that would take a month would require on the order of $10^7$ physical qubits.
There may be other classes of problems that become runtime-limited when fewer or more physical qubits are required.
Thus, in Figure \ref{fig:regimes} we depict the transition from EFTQC to FTQC as the green-to-blue gradient ranging from $10^6$ to $10^8$.

Second, we consider the left-hand side of Equation \ref{eq:success_condition}. Most of the parameters are contained in the factor $A\alpha/\epsilon p_C$. In Section \ref{subsec:eftqc_example} we will explain the importance of this factor in quantifying the ``burden'' placed on the elementary fault-tolerant protocols.
Equation \ref{eq:success_condition} shows that decreasing this burden factor affords a decrease in the number of physical qubits $Q_{\textup{phys}}$.
Alternatively, when fixing the number of physical qubits, a reduction in the burden factor affords an \emph{increase} in the number of logical qubits, and subsequently the maximum problem size or ``reach'' of the quantum computer.
The methods introduced in Section \ref{subsec:eftqc_review} will be understood to reduce this burden factor, enabling algorithms to be run using fewer physical qubits, though at the cost of an increase in runtime.

Figure \ref{fig:scalability} shows the contours of solutions to Equation \ref{eq:success_condition} for several scalability values $s$.
The most striking feature is that, for the finite values of scalability ($s<\infty$), there is a maximum-size instance (measured by $Q_L$) that the architecture can accommodate using the QPE algorithm.
For example, in the case of $s=3.5$,  $p_0 = 0.0001$, and $p_{\textup{th}}=0.01$, we find that the largest instance that can be accommodated (i.e. the ``reach'' of the quantum architecture) is $Q_L\approx 90$.
The maximum number of logical qubits $Q_L^{\textup{max}}$ can be solved for by setting $Q_{\textup{phys}}=Q_{\textup{phys}}^{\textup{opt}}$in Equation \ref{eq:success_condition} and solving for $Q_L$,
\begin{align}
Q_L^{\textup{max}}=\frac{Q_{\textup{phys}}^{\textup{opt}}}{8s^2\beta^2W\left(\sqrt{\left(\frac{A\alpha}{p_{C}}\right)^{\frac{1}{\beta}}\frac{Q_{\textup{phys}}^{\textup{opt}}}{8s^2\beta^2}}\right)^2},
\end{align}
where $W(x)$ is the solution to $W(x)\exp(W(x))=x$, known as the Lambert $W$ function. Using the upper bound of $W(x)\leq \ln(x)$, we can lower bound the maximum qubit number as
\begin{align}
\label{eq:QLmaxbound}
Q_L^{\textup{max}}\geq\frac{\left(\frac{p_{\textup{th}}}{p_0}\right)^s}{2e^2s^2\beta^2\ln\left(\left(\frac{A\alpha}{p_{C}}\right)^{\frac{1}{\beta}}\frac{\left(\frac{p_{\textup{th}}}{p_0}\right)^s}{8e^2s^2\beta^2}\right)^2},
\end{align}
where we have used the expression for $Q_{\textup{phys}}^{\textup{opt}}$.
This maximum solvable problem size motivates the question explored in the next section:
\emph{with a fixed scalability, is it possible to extend the ``reach'' of a quantum architecture using algorithms designed for finite scalability?}

\section{Quantum Algorithms for Early Fault-tolerant Quantum Computers}

\label{sec:eftqc}

\subsection{Review of methods}

\label{subsec:eftqc_review}

\begin{table}[htbp]
\caption{Quantum Algorithms for EFTQC}
\label{tab:eftqc}
\begin{tabular}{|l|l|}
\hline
\textbf{Task} & \textbf{Algorithm} \\ \hline
\multirow{3}{*}{Phase Estimation} 
& Time-Series Estimator \cite{o2022efficient} \\
& Robust Fourier Estimation \cite{kshirsagar2022proving,liang2023modeling} \\
 & Modified Robust Phase Estimation \cite{ni2023low} \\ \hline
\multirow{3}{*}{Multiple-Eigenvalue Estimation} & Adaptive Multi-Order Phase Estimation \cite{dutkiewicz2022heisenberg} \\
& Robust Multiple-Phase Estimation \cite{li2023low} \\
 & Multi-Modal, Multi-Level QCELS \cite{ding2023simultaneous} \\
 & Observable Dynamic Mode Decomposition \cite{shen2023estimating} \\ \hline
\multirow{1}{*}{Ground State Property Estimation} & Heaviside Filter Property Estimation \cite{zhang2022computing} \\ \hline
\multirow{3}{*}{Amplitude Estimation} & $\alpha$-QPE \cite{wang2019accelerated} \\
 & Robust Amplitude Estimation  \cite{wang2021minimizing} \\
 & Power-law AE (also QoPrime AE) \cite{giurgica2022lowa} \\ \hline
\multirow{6}{*}{Ground State Energy Estimation} & 
Quantum Filter Diagonalization \cite{parrish2019quantum} \\
 &
Multireference Selected Quantum Krylov \cite{stair2020multireference} \\
 &
Variational Quantum Phase Estimation \cite{klymko2022real} \\
 &
Fourier Filtering \cite{lin2022heisenberg} \\
 & Statistical Phase Estimation \cite{wan2022randomized} \\
 & Gaussian Filter \cite{wang2022quantum} \\
 & QCELS \cite{ding2022even} \\
 & Quantum Lanczos \cite{kirby2023exact} \\
 & Rejection Sampling \cite{Wang2023} \\
 & Gaussian QPE \cite{rendon2023low} \\ \hline
\multirow{2}{*}{Ground State Preparation} & Gaussian Booster \cite{wang2022state} \\
 & QET-U \cite{dong2022ground}\\
 & Single-Ancilla Lindbladian \cite{ding2023single} \\ 
 \hline
\end{tabular}
\end{table}
The previous section ended with a question about how to extend the reach of quantum computers that have limited scalability, or in other words, early fault-tolerant quantum computers.
A growing body of work in quantum algorithms has developed a suite of methods that might be used to address this problem.
Such algorithms solve the tasks of phase estimation \cite{o2022efficient, kshirsagar2022proving, ni2023low, liang2023modeling}, multiple-eigenvalue estimation \cite{dutkiewicz2022heisenberg, li2023low, ding2023simultaneous, shen2023estimating},
ground state property estimation \cite{zhang2022computing}, amplitude estimation \cite{wang2019accelerated, wang2021minimizing, koh2022foundations, giurgica2022lowa}, ground state energy estimation \cite{lin2022heisenberg, wan2022randomized, wang2022quantum, ding2022even, kirby2023exact, Wang2023}, and ground state preparation \cite{wang2022state, ding2023single}.
In this subsection we provide a non-exhaustive review of the literature in this area.
Note that some of the works do not use the term ``EFTQC''. Nevertheless, we include them because of their influence on later works \cite{wang2019accelerated, wang2021minimizing, giurgica2022lowa} or the similarity in their motivations \cite{shen2023estimating, kirby2023exact}.

These algorithms have typically been developed with certain improvements in mind that include: the reduction of logical qubit number \cite{lin2022heisenberg, ding2023single}, the reduction of number of operations per circuit \cite{wang2019accelerated, wang2021minimizing, wang2022quantum}, the reduction of expensive operations \cite{campbell2021early, wan2022randomized} (e.g. non-Clifford operations like T gates and Toffoli gates), and establishing or increasing the robustness to error \cite{wang2021minimizing, kshirsagar2022proving, li2023low, liang2023modeling, ding2023robust}.
In many cases, achieving these improvements comes at a cost.
The predominant cost is an increase in the number of circuit repetitions (also known as the ``sample complexity'', ``number of samples'' or ``shots''), and, subsequently, runtime.
Another cost is an increase in classical processing (e.g. converting the measurement outcome data from the many circuit repetitions into the estimate of the ground state energy).
In the next subsection we will detail an example algorithm where these trade-offs can be easily understood.

One of the first algorithms suited for early fault-tolerant quantum computers was the so-called $\alpha$-VQE method \cite{wang2019accelerated}. 
This method for solving the task of amplitude estimation enables a trade-off between the number of quantum operations per circuit $O(1/\epsilon^{\alpha})$ and the number of circuit repetitions $\tilde{O}(1/\epsilon^{2(1-\alpha)})$ (where $\tilde{O}$ indicates that we ignore polylog factors), set by a tunable parameter $\alpha$.
Later, Wang et al. introduced a variable-depth amplitude estimation algorithm that is robust to substantial amounts of circuit error \cite{wang2021minimizing}.
Similar methods were explored in the context of quantum algorithms for finance \cite{alcazar2022quantum, giurgica2022lowa} and some have been implemented on quantum hardware \cite{katabarwa2021reducing, giurgica2022lowb}.

Another thread in the development of quantum algorithms for early fault-tolerant quantum computing has focused on problems related to physical systems such as electronic structure or condensed matter systems specified by their Hamiltonians.
In this direction, one of the first papers to introduce the phrase ``early fault-tolerant'' was \cite{campbell2021early}, where the author reduces the counts of expensive non-Clifford operations to make simulation of the Fermi-Hubbard model more amenable to smaller fault-tolerant quantum computers.
Other methods have sought to reduce the logical qubit requirements.
Previous approaches had mostly been based on the quantum phase estimation (QPE) algorithm \cite{Kitaev2003}, which uses additional ancilla qubits for reading out the phase\footnote{Note that the so-called semi-classical Fourier transform can be used to reduce the ancilla count to one, though it requires mid-circuit measurement, reset, and feed-forward at the logical level}. 
A method for estimating the spectrum of a Hamiltonian without the use of QPE (and its ancilla qubit overhead) was introduced in \cite{somma2019quantum}; instead, the spectrum is estimated by classically post-processing measurement outcome data from Hadamard tests of the $c$-$e^{iHt}$ circuit.
Then, one of the first papers to motivate their algorithm development in the context of early fault-tolerant quantum computers was \cite{lin2022heisenberg}.
They developed a novel post-processing technique for the measurement outcome data generated in \cite{somma2019quantum} 
and carried out an analysis of their ground state energy estimation (GSEE) algorithm showing that they could achieve a runtime with Heisenberg-limit scaling of $O(1/\epsilon)$, compared to the $O(1/\epsilon^4)$ runtime of \cite{somma2019quantum} (when improved and applied to the task of GSEE).
Placing their work in the context of early fault-tolerant quantum computing, this work solidified this new research direction in quantum algorithms and helped place earlier works in the context of early fault-tolerant quantum computing.
By combining the insights of \cite{lin2022heisenberg}, linear combination of unitaries \cite{childs2012hamiltonian}, and QDRIFT \cite{campbell2019random}, the authors of
\cite{wan2022randomized} developed a method to 
exploit structure in the Hamiltonian to make the overall complexity of GSEE independent of the number of terms in the Hamiltonian.
They also demonstrate that their method enables trading the number of operations per circuit for number of circuit repetitions.
A methodology similar to the works above has been applied to the task of estimating ground states properties \cite{zhang2022computing}, which is often required in industrially-relevant quantum chemistry calculations \cite{steudtner2023fault, o2022efficient}.

The EFTQC algorithms developed for amplitude estimation \cite{wang2019accelerated, wang2021minimizing, giurgica2022lowa} established a trade-off between operations per circuit and circuit repetitions.
These result in tuning the runtime between Heisenberg limit scaling $O(1/\epsilon)$ and the central limit scaling $O(1/\epsilon^2)$.
This raises the question of whether a similar trade-off can be established for the task of ground state energy estimation.
Such circuit trading was established in \cite{wang2022quantum}. They showed an exponential reduction in the number of operations per circuit in terms of accuracy dependence (i.e. a reduction from $\tilde{O}(1/\epsilon)$ to $\tilde{O}(\log{1/\epsilon})$).
This reduction in number of operations per circuit comes at the cost of an increase in circuit repetitions from $\tilde{O}(\log{1/\epsilon})$ to $\tilde{O}(1/\epsilon^2)$. 
With this method, the minimal number of operations per circuit is $\tilde{O}(1/\Delta)$, where $\Delta$ is a lower bound on the spectral gap of $H$.
Often it is the case that $\Delta$ is larger than $\epsilon$, enabling a reduction in number of operations using this method~\cite{wang2022quantum}.
Later, Ding and Lin \cite{ding2022even} established a similar result using an approach based on numerically fitting a parameterized curve to a set of estimated expectation values, which they refer to as quantum complex exponential least squares (QCELS).
They also showed that, assuming the ground state overlap $|\bra{\psi}gs\rangle|=\gamma$ is sufficiently close to 1, the circuit depth of energy estimation can be made arbitrarily small while still retaining the Heisenberg limit scaling.

These two methods that enabled a reduction in number of operations per circuit \cite{wang2022quantum, ding2022even} required a runtime and number of circuit repetitions scaling as $O(1/\gamma^4)$.
In contrast, for methods using more operations per circuit, a runtime scaling of $O(1/\gamma^2)$ \cite{knill2007optimal} (and even $O(1/\gamma)$ \cite{dong2022ground}) was shown to be possible.
This motivated the search for algorithms that could improve the runtime scaling with respect to overlap to $O(1/\gamma^2)$, while also using few operations per circuit.
The first work to achieve this was \cite{Wang2023}. 
This algorithm uses the quantum computer in a very different manner compared to previous approaches.
To estimate the ground state energy, a classical computer first generates a uniformly random sampling of energy values on an interval expected to contain the ground state energy.
Then, for each sample a quantum circuit is designed such that a binary measurement outcome from the quantum computer is used to decide whether or not that sample should be accepted or rejected.
The set of accepted samples are proven to be drawn from a Gaussian distribution centered about the ground state energy and the mean of these samples will be close to this value.
The number of operations per circuit can be tuned anywhere from $O(1/\epsilon)$ to $O(1/\Delta)$, where the consequence is a broadening of the Gaussian peak width (requiring more samples to achieve the same accuracy).
Later, \cite{rendon2023low}  also developed a ground state energy estimation algorithm with $O(1/\Delta)$ operations per circuit and $O(1/\gamma^2)$ circuit repetitions based on a Gaussian-filter variant of quantum phase estimation.
Although this uses more ancilla qubits like QPE, \cite{rendon2023low} shows that the number of operations per circuit is reduced compared to previous methods. 

Another thread of research in ground state energy estimation has drawn on methods from numerical linear algebra to classically postprocess quantum measurement outcome data in a more efficient and robust manner \cite{parrish2019quantum, stair2020multireference, klymko2022real, kirby2023exact, stair2023stochastic, shen2023estimating}.
These methods employ techniques like filter diagonalization \cite{parrish2019quantum, klymko2022real}, Lanczos methods \cite{kirby2023exact}, and dynamic mode decomposition \cite{shen2023estimating}.
Some of these methods \cite{stair2023stochastic, kirby2023exact} have been shown to only require a number of operations per circuit scaling as $\tilde{O}(1/\Delta)$, similar to \cite{wang2022quantum}.
However, in the case of \cite{kirby2023exact}, the runtime upper bound has a scaling of $\tilde{O}(1/\Delta^2)$, compared to the $\tilde{O}(\Delta)$ runtime scaling of \cite{wang2022quantum}.
An important direction for future work will be to carry out empirical studies that give more realistic estimates for the runtimes, required operations per circuit, and robustness of these algorithms.

As mentioned above, all ground state energy estimation methods have a runtime that depends on the overlap between the input trial state and the ground state $\bra{\psi}gs\rangle=\gamma$.
The runtime of these ground state energy estimation algorithms can be improved by using a ground state preparation method to improve this overlap before running the ground state energy estimation algorithm. 
For papers analyzing the interplay between ground state energy estimation and state preparation see \cite{pathak2023quantifying, gratsea2022reject}.
An excellent overview of ground state preparation algorithms is presented in Table II of \cite{dong2022ground}.
In the regime of early fault-tolerant quantum computing, it may be advantageous to use ground state preparation methods that reduce the number of operations per circuit.
One such method was proposed in 
\cite{wang2022state}, where an approximate Gaussian filter of varying width can be used to suppress high energy states and boost the overlap with the ground state.
More recently, \cite{ding2023single} introduced a novel ground state preparation method based on Lindblad dynamics, engineering a process that has the ground state as the unique steady state.

Finally, we discuss another important thread of research for early fault-tolerant quantum computing: \emph{robustness}.
As shown in Section \ref{subsec:qpe}, $p_C$, the circuit error rate that the algorithm can tolerate, plays a role in determining the fault-tolerant overhead.
An increase in the robustness (i.e. $p_C$) reduces the fault-tolerant overhead.
A canonical reference on the robustness of quantum algorithms is \cite{kimmel2015robust}, which introduced the robust phase estimation algorithm.
They showed that their variant of quantum phase estimation was able to tolerate a substantial circuit error rate of $\sim 35 \%$.
Such robustness analysis is especially important when reduction of fault-tolerant overhead is essential, as in quantum algorithms suited for early fault-tolerant quantum computers.

One of the first works to analyze the robustness of a quantum algorithm in the EFTQC setting was \cite{kshirsagar2022proving}.
Here, the simple robust phase estimation algorithm was introduced and its robustness was analyzed with respect to two different models of algorithmic noise (i.e. a model of how the measurement outcome probabilities are impacted).
See Section \ref{subsec:eftqc_example} for a discussion of these algorithmic noise models.
A variant of the robust phase estimation algorithm was developed in \cite{liang2023modeling} that enables circuit trading.
Here the robustness of the algorithm is analyzed with respect to the exponential decay model (see Section \ref{subsec:eftqc_example}).
Most recently, \cite{ding2023robust} developed an algorithm for ground state energy estimation that is provably robust with respect to the exponential decay model (see Section \ref{eq:exp_decay}).

The works presented above represent a foundation for an increasingly important research direction: the development of quantum algorithms suited to the capabilities of finite-scalability quantum computers.
They are built to reduce logical qubit number, to enable a trade-off between gate count and sample cost, and they are robust to error in the circuit.
All of these features contribute to reducing the burden placed on fault-tolerant operations and thus reducing fault-tolerant overheads.
This helps to run larger problem instances on earlier quantum computers, or, in other words to ``extend the reach'' of a finite-scalability quantum architecture.
In the following subsection we will make these concepts more clear through an example.

\subsection{Example: randomized Fourier estimation under finite scalability}
\label{subsec:eftqc_example}

Section \ref{subsec:qpe} ended with the question of how we might extend the reach of finite scalability quantum computers.
The previous subsection overviewed a host of quantum algorithms suited for addressing this question.
In this section we take one quantum algorithm from the previous section and quantitatively investigate its ability to extend the reach of a finite scalability quantum computer for the task of phase estimation.

We use as our example the randomized Fourier estimation (RFE) algorithm as introduced in \cite{kshirsagar2022proving} and adapted for trading circuit repetitions for number of operations per circuit in \cite{liang2023modeling}.
The RFE algorithm solves the task of phase estimation introduced in Section \ref{subsec:qpe}. 
It is an alternative to the standard quantum phase estimation (QPE) algorithm \cite{NC_book} and related algorithms such as robust phase estimation (RPE) \cite{kimmel2015robust}.

We consider the RFE algorithm to be a prototypical quantum algorithm suited for early fault-tolerant quantum computing given that it has the following features:
\begin{itemize}
    \item \textbf{Qubit conservation}: the (high-level) circuit conserves qubit count by using just one ancilla qubit.
    \item \textbf{Circuit trading}: the number of operations per circuit is tuned by input parameter $K$, enabling a trade-off between this quantity and the required number of circuit repetitions.
    \item \textbf{Robustness}: the algorithm is robust to circuit error and this robustness can be understood in terms of a signal corrupted by a noise floor.
\end{itemize}
As we will show, and like many of the other EFTQC algorithms introduced in Section \ref{subsec:eftqc_review}, these features equip the algorithm to accommodate limited scalability in the early fault-tolerant quantum computing regime.
Furthermore, RFE is very simple, helping to facilitate discussion of these algorithmic concepts relevant to early fault-tolerant quantum computing.

We will a) briefly review randomized Fourier estimation, and then investigate b) how trading circuit repetitions for decrease of operations and c) how robustness to error help to increase the problem instance size (i.e. the ``reach'') that can be solved with a finite scalability architecture.

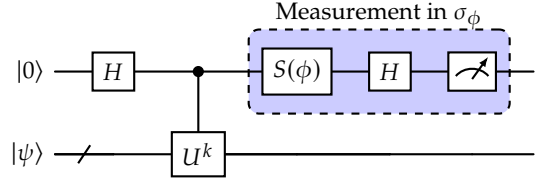
\begin{figure}[!htb]
\centering
\begin{quantikz}
\lstick{$\ket{0}$} &  \gate[wires=1][0.1cm]{H}  &
\ctrl{1} & \gate[wires=1][0.1cm]{S(\phi)}\gategroup[wires=1,steps =3,style={dashed , rounded corners,fill=blue!20, inner xsep=2pt}, background]{Measurement in $\sigma_{\phi}$}&\gate[wires=1][0.1cm]{H} & \meter{} & 
\\
\lstick{$\ket{\psi}$} & \qw \qwbundle{}  & \gate{U^k} & \qw & \qw & &
\end{quantikz}
\caption{An archetypal circuit template used by many EFTQC algorithms. The measurement outcome probabilites depend on $\ket{\psi}$ and $U$ as $\textup{Pr}(\pm 1|k,\phi)=\frac{1}{2}(1\pm\cos(k\theta+\phi))$. Measurement outcomes can be processed to
In the case of the Randomized Fourier Estimation (RFE) algorithm, the measurement outcomes encode the
The parameter $k$ is uniformly randomly chosen among $\{0, \ldots, K-1\}$ for each circuit repetition. $K$ then controls the maximal circuit depth and is used to reduce the number of operations per circuit.
The boxed-up elements in blue can be collectively interpreted as a measurement with respect to the observable $\sigma_{\phi}=\cos(\phi)\sigma_x-\sin(\phi)\sigma_y$, where $\sigma_x$ and $\sigma_y$ are the conventional Pauli operators and $S(\phi)=\begin{bmatrix}
1 & 0\\
0 & \exp(i\phi)
\end{bmatrix}$.
}
\label{fig:hadamard_test}
\end{figure} 

\paragraph{RFE Intro} The RFE algorithm relies on the Hadamard test circuit (as depicted in Figure \ref{fig:hadamard_test}). Each Hadamard test circuit is parameterized by the circuit depth ($k$) and a phase parameter ($\phi$).
The output measurement probabilities correspond to an oscillatory function that encodes $\theta$:
\begin{align}
\textup{Pr}(z|\theta;k,\phi)&=\frac{1}{2}(1+z\textup{Re}(e^{i\phi}\bra{\psi}U^k\ket{\psi}))\\
&=\frac{1}{2}(1+z\cos(k\theta+\phi))\\
\end{align}
It is convenient to view the expected value of $z$, which is $g(k)=\cos(k\theta+\phi)$, as the true signal encoding $\theta$.
The phase $\theta$ is then estimated from measurement outcome data in a manner similar to estimating the frequency of a noisy estimate of $g(k)$. The parameters $k$ and $\phi$ are chosen uniformly randomly in each sample, with $k\in[0,K-1]$ and $\phi$ ranging between 0 and $2\pi$. Each measurement outcome $z$ obtained from the circuit is used to form an unbiased estimator $\hat{f}_j=2ze^{-i2\pi kj/J}e^{-i\phi}$ of the discrete Fourier transform of the signal $g(k)$, where $J$ is an algorithm parameter that sets the grid size of the Fourier spectrum. 
The estimate of the Fourier signal can be made more accurate by taking multiple samples and averaging them:
\begin{align}
    \hat{f}_j=\frac{1}{M}\sum_{i=1}^{M} \hat{f}_j^{(i)}.
\end{align}
By accumulating enough measurement outcomes, one can estimate $\theta$ accurately (i.e. within $\epsilon$) with high probability (i.e. less than $1-\delta$) by locating the tallest peak (or point of largest magnitude) in the estimate of the discrete Fourier transform,
\begin{align}
\hat{\theta} = \frac{2\pi}{J}\textup{argmax}_j \frac{1}{M}\sum_{i=1}^{M} \hat{f}_j^{(i)}.
\end{align}
The algorithm's accuracy is limited by parameter $J$, which is set to ensure that the Fourier resolution matches the desired accuracy. 

\[\begin{array}{ll}
\textbf{Algorithm Parameters} \\
    J  & \text{Sets the Fourier domain grid spacing.} \\
    K  & \text{Sets the maximum number of } c\textup{-}U \text{ per circuit.} \\
    M  & \text{Sets the number of circuit repetitions (i.e. samples).} \\
    [2em]    
\textbf{Error and Confidence Requirements} \\
    \epsilon = 2\pi/J  & \text{Ensures that the $\theta$-adjacent discrete frequencies are accurate.} \\
    \delta = 8J\exp(-M/W(K, J, \lambda))  & \text{Ensures that enough samples are taken, given $K$, $J$, and $\lambda$ (see below).} \\[2em]
\textbf{Operations Per Circuit} \\
\mathbb{E}G_C = \frac{K-1}{2}G_U   & \text{Expected value, with max being } (K-1)G_U \\[2em]
\textbf{Circuit Repetitions} \\
M = W(K,J,\lambda)\log(\frac{16\pi}{\delta\epsilon}) &  \text{Number of samples needed. See \cite{liang2023modeling} for definition of }W(K,J,\lambda). \\[2em]
\end{array}
\]

\begin{figure}[ht]
    \centering
    \adjustbox{cfbox=black 0.25pt}{\includegraphics[width=0.8\textwidth]{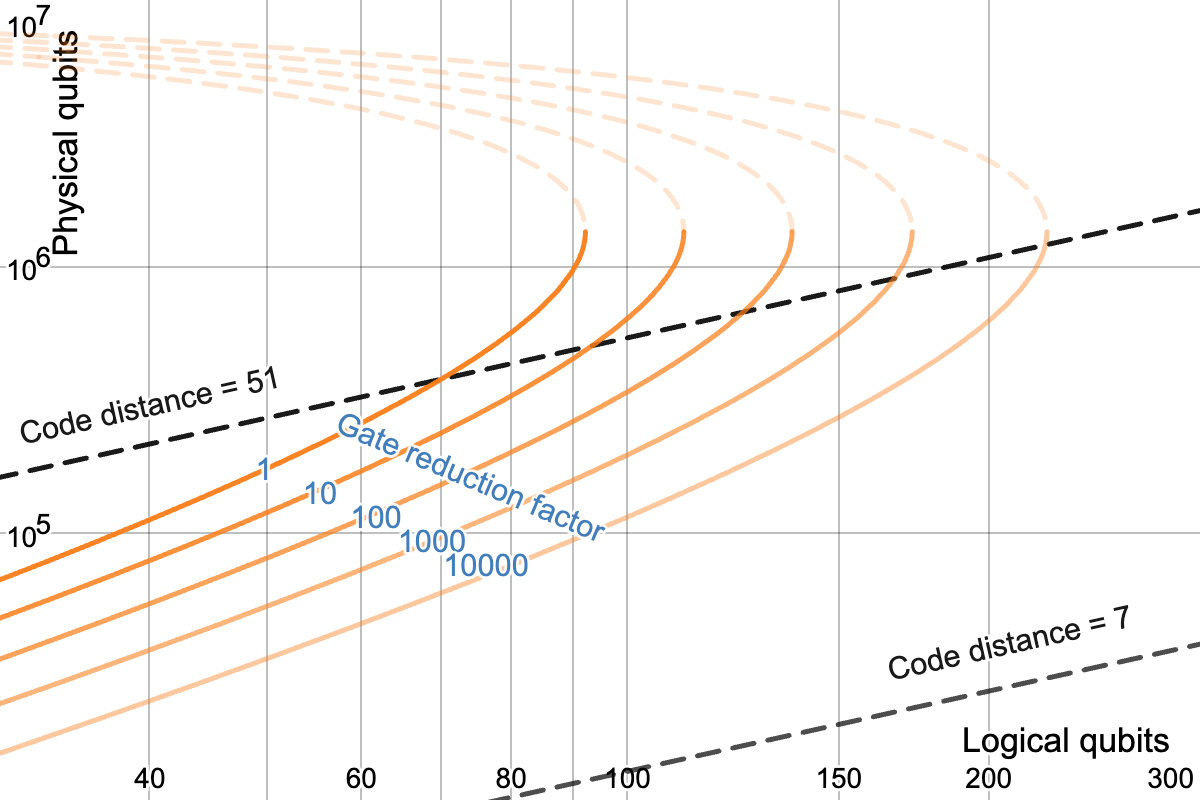}}
    \caption{This plot shows that, under the scalability model, the EFTQC algorithm robust Fourier estimation (RFE) can extend the reach of the quantum computation from 90 logical qubits to over 200 logical qubits.
    This is achieved by either reducing the number of $c$-$U$ used per circuit or increasing the tolerable circuit error rate $p_C$ 
    in the RFE algorithm.
    Both of these reduce the burden factor $A\alpha/\epsilon p_C$ appearing in Equation \ref{eq:success_condition}.
    This increase in the ``reach'' of the quantum computer comes at the cost of an increase in the runtime (roughly by the burden factor), which is a combination of the decrease in time per circuit and increase in number of circuit repetitions.
    Here we take the scalability to be $s=3.5$ with $p_0=10^{-4}$, which implies that the optimal number of physical qubits is $Q_{\textup{phys}}^{\textup{opt}}=\frac{1}{e^2}\left(\frac{p_{\textup{th}}}{p_0}\right)^s\approx1.35\times 10^6$. An editable version of the plot can be accessed here: \href{https://www.desmos.com/calculator/nf43nafwet}{https://www.desmos.com/calculator/nf43nafwet}}
    \label{fig:rfe}
\end{figure}

\paragraph{Circuit trading}
We now describe how this algorithm is able to trade number of operations per circuit for circuit repetitions. 
The maximum number of operations per circuit (in expectation) is $(K-1)G_U$, where $G_U$ is the number of operations in a single $c$-$U$.
In the quantum phase estimation algorithm, $1/\epsilon$ calls are made to $c$-$U$,
corresponding to setting $K \approx 1/\epsilon$.
In RFE we can reduce the number of operations per circuit by setting $K$ to any value less than $1/\epsilon$.
This reduction in $K$ reduces the burden factor in Equations \ref{eq:success_condition} and \ref{eq:QLmaxbound} proportionally. 
Figure \ref{fig:rfe} shows how varying reductions in the burden factor lead to an increase in the problem size that RFE can accommodate.
Equation \ref{eq:QLmaxbound} predicts that this increase in problem size grows as $O(1/\ln^2(B))$ with burden factor $B$.
For the specific example considered, the largest problem instance can be increased from $90$ to over $200$ by decreasing $K$ by a factor of 100,000.

As mentioned previously, circuit trading means that a decrease in operations per circuit comes at the cost of an increase in the number of circuit repetitions.
This trade-off can be understood as follows.
Decreasing $K$ causes the \emph{width} of the peak in the discrete Fourier spectrum to increase.
With the spectrum being more flat near the peak, smaller amounts of noise in the signal are able to shift the peak location more than $\epsilon$ (leading to algorithm failure).
This statistical sampling noise must then be reduced by taking more samples.
The analytic relationship is given in the appendix of \cite{liang2023modeling}.
This describes the nature of the trade-off between operations per circuit and circuit repetitions.

\paragraph{Robustness}
The RFE algorithm has been analyzed in previous work with respect to three different algorithmic noise models: adversarial noise and Gaussian noise \cite{kshirsagar2022proving} and exponential decay noise \cite{liang2023modeling}.
We give brief explanations of how the Gaussian noise and the exponential decay noise impact the algorithm performance and thus explain the robustness of the RFE algorithm to a particular model of noise.
In \cite{kshirsagar2022proving}, the Gaussian noise model is analyzed, wherein it is assumed that, for each circuit (labeled by $k$), the output probability has been corrupted by a small perturbation drawn from a Gaussian distribution,
\begin{align}
\textup{Pr}(z|\theta;k,\phi)
&=\frac{1}{2}(1+z(\cos(k\theta+\phi)+\eta_k)),
\end{align}
where each $\eta_k$ has been drawn from a Gaussian distribution with mean zero and standard deviation $\sigma$.
How does this impact the performance of the algorithm?
The $\eta_k$ can be understood to corrupt the expected value of $z$ (i.e. the signal $g(k)$).
This impacts the Fourier spectrum by adding a ``noise floor'' related to the Fourier transform of the $\eta_k$.
The algorithm can still succeed as long as this noise floor does not shift the location of the peak by more than $\epsilon$. The authors of \cite{kshirsagar2022proving} proved that if $\sigma$ is below a certain quantity (dependent on $\epsilon$ and $\delta$) then the algorithm can succeed with more than $1-\delta$ probability.

In \cite{liang2023modeling}, the exponential decay model is derived from a lower-level noise model.
The exponential decay model assumes that the likelihood function now includes a factor that decreases exponentially in $k$,
\begin{align}
\label{eq:exp_decay}
\textup{Pr}(z|\theta;k,\phi)
&=\frac{1}{2}(1+ze^{-k\lambda}\cos(k\theta+\phi)),
\end{align}
with decay parameter $\lambda$.
Experiments \cite{katabarwa2021reducing, giurgica2022lowb} 
show that this model is accurate for small systems.
This exponential decay factor causes the expected value of $z$ (i.e. the underlying signal $g(k)$) to attenuate as $k$ is increased.
In the Fourier domain, this attenuation translates into an attenuation of the peak (see \href{https://www.desmos.com/calculator/7fwvz6fxgb}{this desmos plot}).
As with the peak broadening due to reducing $K$, a smaller amount of statistical noise is sufficient to shift the location of the estimated peak more than $\epsilon$.
Accordingly, more samples must be taken to sufficiently reduce this statistical noise. 

Under the assumption that the exponential decay model holds exactly, \cite{liang2023modeling} shows that with arbitrarily large decay parameter $\lambda$, the algorithm can generate an $\epsilon$ accurate estimate with probability greater than $1-\delta$.
In other words, the algorithm can be made arbitrarily robust.
The reason is that the exponential decay error does not shift the location of the peak in the Fourier spectrum of the expected signal.
This increase in robustness translates into a decrease in the burden: allowing the circuit error rate $p_C$ to increase towards 1 increases the allowed logical error rate $p_L$, decreasing the burden factor.

Consider a reduction in the burden on account of an increase in the tolerable circuit error rate $p_C$, which quantifies the the robustness of the algorithm.
Note that in the case where $p_C$ is close to $1$, a better approximation than the union bound can be used to replace $p_C$ with $\ln (1/(1-p_C))$, which grows to infinity as $p_C\rightarrow \infty$.
We remark that in the case of the exponential decay model, the circuit error rate is $p_C = 1-e^{-k\lambda}$, which leads to $\ln (1/(1-p_C))=k\lambda$.
Therefore, as we allow for an increase in $\lambda$, the burden factor is reduced proportionally (where we keep in mind that, for small values of $p_C$, the burden factor scales proportionally to it).

We previously discussed Figure \ref{fig:rfe} in the context of circuit trading.
This figure can also be used to demonstrate the impact of increased robustness.
Considering an increase in $p_C$ to be the cause of the burden factor reduction, Figure \ref{fig:rfe} shows how the reach of the quantum computer is increased accordingly.
As with circuit trading, there is a price paid for this extended reach of the quantum computer: for the RFE algorithm, \cite{liang2023modeling} shows that the runtime grows exponentially in $\lambda$ for $\lambda\geq 1/2$ (where $K$ is set to its minimum value of $2$).
Therefore, in practice, there may be an upper limit to the degree of robustness, beyond which the runtime becomes too large to be practical.
This is an issue that many error mitigation techniques face \cite{cai2022quantum}.
This similarity may not be surprising in that the way RFE accommodates error is a type of error mitigation.

In practice, the exponential decay model is not exact.
Instead, we expect that in any given device and compilation of $c$-$U$, the likelihood function will include some deviation (possibly varying over time) from the exponential decay model likelihoods.
While in the exact exponential decay model the Fourier peak location is unchanged, allowing deviations from this model can shift the location of the peak.
This sets a lower limit to the achievable accuracy $\epsilon$, a feature which is found in the bounded adversarial noise model and the Gaussian noise model of \cite{kshirsagar2022proving}.

We have demonstrated how the RFE algorithm, as an archetypal EFTQC algorithm, enables a reduction in the burden placed on the fault-tolerant protocols.
Figure \ref{fig:rfe} demonstrates how larger problem instance sizes can be accommodated by either reducing the number of operations per circuit (decreasing $K$) or by increasing the robustness of the algorithm (increasing $p_C$). 
This is because the burden factor $\frac{A\alpha}{p_C}$ incorporates both of these quantities.
For both examples of reducing the burden factor, there is an increase in the runtime of the algorithm.
Although the RFE algorithm enables parallelizing the circuit repetitions over multiple quantum computers to reduce runtime, the runtime is expected to be a bottleneck for many applications.
Therefore, the runtime costs of reducing the fault-tolerance burden must be carefully considered.
See \cite{liang2023modeling} for an quantitative account of such runtime costs for RFE.
We leave a thorough investigation of the runtime costs of decreasing the fault-tolerance burden for EFTQC algorithms to future work.

\section{Discussion and Outlook}

\label{sec:discussion}


In this perspective we investigated the regime between NISQ and FTQC, which is referred to as ``early fault-tolerant quantum computing''.
To understand the prospects for utility in this regime,
we proposed a simple computational model to quantitatively capture the performance of quantum architectures within these three regimes.
The \emph{scalability} model characterizes the ability of a quantum hardware vendor to provide systems with low physical error rates as the requested number of physical qubits is increased.
This differs from previous approaches that assume a scale-independent performance for their quantum architectures \cite{kim2022fault, goings2022reliably, beverland2022assessing}.
We demonstrated that the QPE algorithm~\cite{cleve1998quantum} compiled to the surface code~\cite{fowler2018low} has a limit on the problem size that can be accommodated 
by a vendor with finite scalability, according to our model.
Unsurprisingly, this is due to scale-dependent error rates (Equation~\ref{eq:scalability}) combined with the diminishing returns of fault-tolerant protocols as the error rates of the device approach the numerically-estimated threshold value \cite{wang2011surface}.
Next, we showed 
that by using an algorithm suited to finite-scalability (the randomized Fourier estimation algorithm~\cite{kshirsagar2022proving}), when granted the same scalability,
the problem size limit can be extended from around 90 qubits (for QPE) to around 130 qubits (using the same number of physical qubits).
This comes at the cost of roughly a 100 times increase in runtime.

The scalability model enabled us to quantitatively discuss the transition from NISQ to EFTQC to FTQC.
At the end of Section \ref{subsec:scalability} we described how the nature of the transition from the regime of NISQ to EFTQC is difficult to predict; future advances might allow for implementing certain fault-tolerant components far sooner than current methods would enable.
However, we mentioned some of the technical 
considerations that might govern the transition and, accordingly, depict this transition in Figure \ref{fig:regimes} to occur through the range of $Q_\textup{phys}^{\textup{max}}$ being 100 to 10,000.
Regarding the transition from EFTQC to FTQC, we described in Section \ref{subsec:scalability} how each regime might be characterized by different bottlenecks;
EFTQC is characterized by the largest solvable problem instances being bottle-necked by the number of available physical qubits (or, better, $Q_{\textup{phys}}^{\textup{opt}}$), whereas FTQC is characterized by the largest solvable problem instances being bottle-necked by runtime.
Accordingly, we explain how this transition might occur in the range of $Q_\textup{phys}^{\textup{max}}$ being $10^6$ to $10^8$.

Different factors, such as hardware, algorithmic, and fault-tolerance advances, play a dominant role in characterizing the EFTQC regime. The recent work of~\cite{Kim2023evidence} provides evidence for the utility of noisy quantum devices in the pre-fault tolerant era and emphasizes the role of hardware advances to achieve this. Moreover, many works have highlighted the importance that quantum algorithm development has in leveraging the capabilities of the quantum devices to their maximum potential~\cite{wang2019accelerated, wang2021minimizing, katabarwa2021reducing, Xiao2023robust}. Recent works have also explored the effect of noise in the performance of quantum algorithms and highlighted the need to use QEC prudently~\cite{alcazar2022quantum, johnson2022reducing, liang2023modeling}. This work is a first attempt to incorporate all the aforementioned factors (hardware, algorithmic, and fault-tolerance advances) in order to validate the assumption that there is a meaningful regime of early fault-tolerant quantum computing methods, which is usually assumed in papers on the subject~\cite{lin2022heisenberg, wang2023quantum, ding2022even}.
What remains to be determined is how rapidly quantum hardware will progress through this regime; or, in other words,
it remains to determine how the scalability of quantum hardware vendors will increase over time.

To put these results into context, recent resource estimates on a variety of molecules relevant to Li-on electrolyte chemistry~\cite{Kim2022Fault-tolerant} show that above $100$ logical qubits would be necessary to tackle such systems. This indicates that extending the reach of the problem size limit from $90$ logical qubits to over $130$ with the framework discussed here might have interesting implications, i.e. allowing to study problems of interest before the realization of FTQC regime. Our results suggest that the EFTQC regime could exist in a meaningful way, i.e. using the same quantum resources compared to FTQC (number of physical qubits and scalability model), while affording the use of a larger number of logical qubits (Fig.~\ref{fig:rfe}). 


This work explored the usefulness of the EFTQC regime for a specific quantum algorithm and QEC model, namely the RFE~\cite{kshirsagar2022proving} and the surface code~\cite{fowler2018low}. The underlying methodology, however, can be easily extended to other algorithms and fault-tolerant protocols while using the suggested or alternative scalability models. Although the proposed model of scalability is quite general, we do not expect it to perfectly fit the scalability profile of vendors over many orders of magnitude. 
But, we anticipate that it could capture the qualitative behavior over at least a few orders of magnitude. 
Moreover, we showed in Appendix \ref{app:log_model} that, even when a more optimistic model is used (specifically a logarithmic model), the qualitative finding remains: there is an upper limit on the size of the quantum computation.

Future work could explore other models of scalability that, for example, might be given directly by the hardware provider and accommodate the features of the architecture as it is scaled. 
Another interesting direction is to adapt the scalability model to address the interplay between quantum error mitigation and quantum error correction~\cite{Priveteau2021ErrorMitigation, Suzuki2022Qauntum} which will help drive the transition from NISQ to EFTQC. Moreover, the proposed framework could be applied to other combinations of algorithms and quantum error correcting codes and be used to examine the utility of the EFTQC regime for other potential application fields of quantum computing.


Our work provides evidence for the utility of the EFTQC regime within a framework that includes crucial factors of quantum computing, such as hardware, algorithm, and fault-tolerance advances. To incorporate the hardware advances, we have introduced a simple scalability model to capture the performance of devices that are continually improving. As it is yet unclear how exactly quantum devices will scale up to incorporate millions or billions of physical qubits~\cite{preskill2018quantum}, the proposed model of scalability is just a first attempt to bridge the gap between NISQ and FTQC. Future works in these directions could help move beyond the NISQ-FTQC dichotomy and further explore how EFTQC might deliver practical quantum advantage at scale.

\section*{Acknowledgements}
\label{sec:ack}
We extend our gratitude to Lin Lin and Zhiyan Ding for their invaluable discussions and for sharing their insights on early fault-tolerant quantum computing. Their substantial feedback significantly enhanced the clarity and presentation of our manuscript.
We thank Shangjie Guo for a careful reading of the manuscript, improving the presentation.
We also thank Daniel Stilck Fran\c{c}a and Guoming Wang for many inspiring conversations on the topic of early fault-tolerant quantum computing, with special thanks to Daniel Stilck Fran\c{c}a for contributing material to an earlier version of the manuscript.
Our appreciation also goes to the team of Zapata scientists whose thoughtful suggestions shaped the presentation of this work. 
We are particularly indebted to Max Radin, Nicole Bellonzi, Vlad Vargas-Calderón, Shima Bab Hadiashar, Matt Kowalsky, Yudong Cao, and others for their constructive critiques and expertise that greatly contributed to the refinement of our ideas and methodologies presented in this paper.
Finally, we thank Al\'{a}n Aspuru-Guzik, Alejandro Perdomo-Ortiz, Aram Harrow and Will Oliver for early discussions on the topic of early fault-tolerant quantum computing that helped motivate this work.

This work was performed while K.G. was a research intern at Zapata AI, Inc. Part of this research was performed while K.G. was visiting the Institute for Pure and Applied Mathematics (IPAM), which is supported by the National Science Foundation (Grant No. DMS-1925919). K.G. acknowledges support from the European Union's Horizon 2020 research and innovation programme under the Marie Skłodowska-Curie grant agreement No. 847517. 

ICFO group acknowledges support from ERC AdG NOQIA; Ministerio de Ciencia y Innovation Agencia Estatal de Investigaciones (PGC2018-097027-B-I00/10.13039/501100011033, CEX2019-000910-S/10.13039/501100011033, Plan National FIDEUA PID2019-106901GB-I00, FPI, QUANTERA MAQS PCI2019-111828-2, QUANTERA DYNAMITE PCI2022-132919,  Proyectos de I+D+I “Retos Colaboración” QUSPIN RTC2019-007196-7); MICIIN with funding from the European Union NextGenerationEU (PRTR-C17.I1) and by Generalitat de Catalunya;  Fundació Cellex; Fundació Mir-Puig; Generalitat de Catalunya (European Social Fund FEDER and CERCA program, AGAUR Grant No. 2021 SGR 01452, QuantumCAT \ U16-011424, co-funded by ERDF Operational Program of Catalonia 2014-2020); Barcelona Supercomputing Center MareNostrum (FI-2023-1-0013); EU (PASQuanS2.1, 101113690); EU Horizon 2020 FET-OPEN OPTOlogic (Grant No 899794); EU Horizon Europe Program (Grant Agreement 101080086 — NeQST), National Science Centre, Poland (Symfonia Grant No. 2016/20/W/ST4/00314); ICFO Internal “QuantumGaudi” project; European Union’s Horizon 2020 research and innovation program under the Marie-Skłodowska-Curie grant agreement No 101029393 (STREDCH) and No 847648  (“La Caixa” Junior Leaders fellowships ID100010434: LCF/BQ/PI19/11690013, LCF/BQ/PI20/11760031,  LCF/BQ/PR20/11770012, LCF/BQ/PR21/11840013). Views and opinions expressed are, however, those of the authors only and do not necessarily reflect those of the European Union, European Commission, European Climate, Infrastructure and Environment Executive Agency (CINEA), nor any other granting authority.  Neither the European Union nor any granting authority can be held responsible for them.

\appendix

\section{Scalability of today's devices}\label{app:ibm}

\begin{figure}[h]
    \centering
    \includegraphics[width=0.6\textwidth]{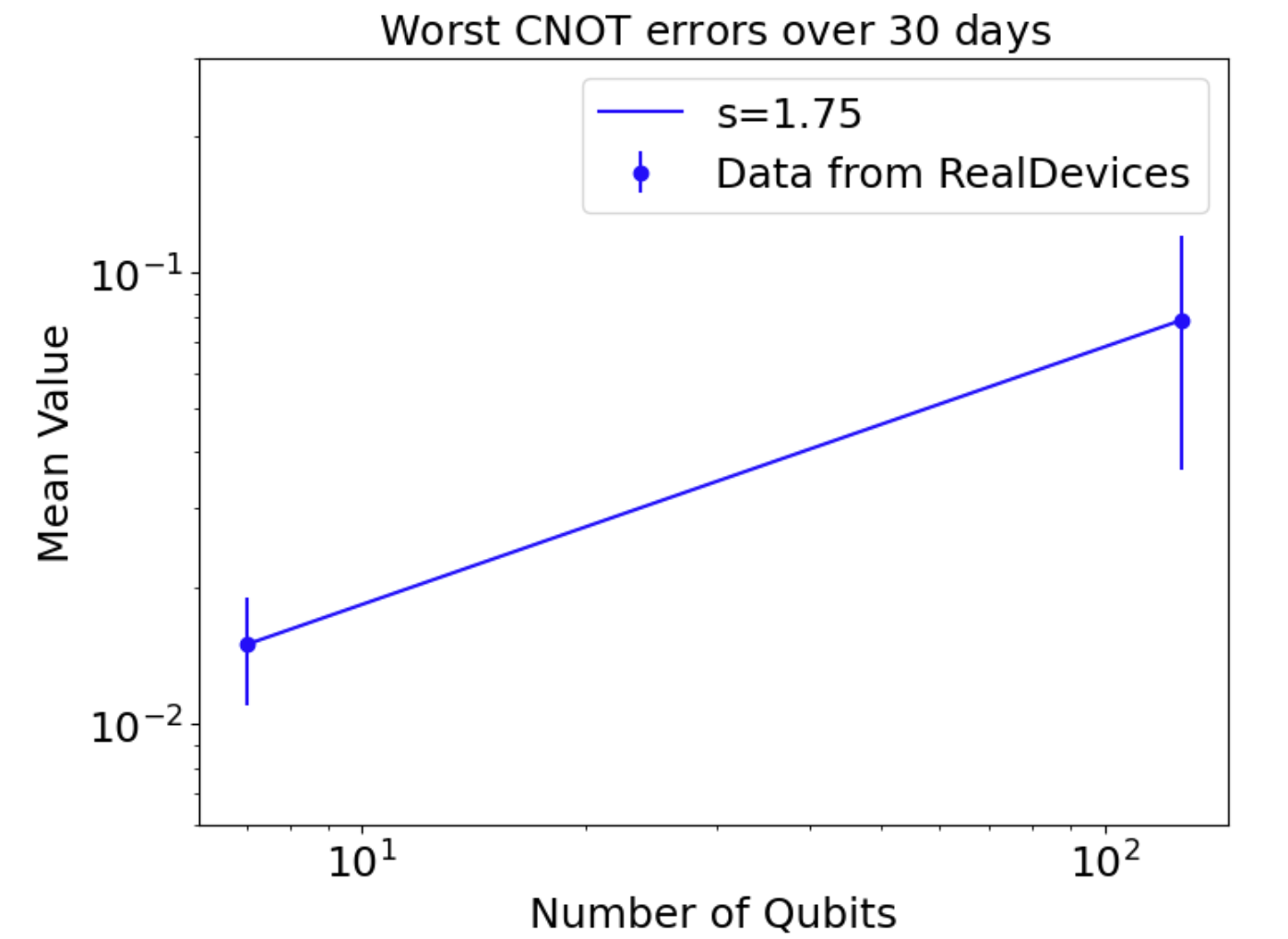}
    \caption{We plot the worst two qubit gate error of two IBM quantum devices on the cloud as a function of the number of qubits. The power law fit (blue line) suggests today's scalability is $s=1.75$ and $p_0 = 0.005$.}
    \label{fig:IBM_scalability}
\end{figure}

Here we estimate the scalability of today's quantum devices. 
To this end, we collected data from two IBM devices available on the cloud, namely $\textit{lagos}$ and $\textit{brisbane}$.
We collected the CNOT error rates at $10$:$00$ am each day over 30 days. 
In Fig.~\ref{fig:IBM_scalability}, we plot the mean and standard deviation of the worst-case CNOT error rates for the studied devices. 
We then make a power-law fit of the data to estimate the $p_0$ and $s$ as introduced in Eq.~\ref{eq:scalability}. We find that $p_0 = 0.005$ and $s=1.75$, which we refer to as today's scalability. Figure \ref{fig:regimes} shows this point to lie in the NISQ regime, despite $p_0$ being below threshold.

\section{Logarithmic scalability model}
\label{app:log_model}

In this section we investigate the implications of a more optimistic scalability model. Instead of the power law model of Equation \ref{eq:scalability}, we consider a logarithmic model for the scalability profile,
\begin{align}
    \label{eq:logscalability}
    p_{\textup{phys}}(Q_{\textup{phys}};\mathcal{V})=p_0\left(1+\frac{1}{\sigma}\ln(Q_{\textup{phys}})\right),
\end{align}
where $\sigma$ is the scalability parameter analogous to $s$ in Equation \ref{eq:scalability}.
The parameterization is chosen such that at $Q_{\textup{phys}}=1$, the function value and slope match those of Equation \ref{eq:scalability}.
With this alternative scalability model, the physical qubit number at which the physical error rate exceeds the threshold value is now
\begin{align}
Q_{\textup{phys}}^{\textup{max}}=\exp\left(\sigma\frac{p_0-p_{\textup{th}}}{p_{\textup{th}}}\right),    
\end{align}
which, compared to the power law model, grows exponentially in the gap between $p_0$ and $p_{\textup{th}}$.

\begin{figure}[ht]
    \centering
    \adjustbox{cfbox=black 0.25pt}{\includegraphics[width=0.6\textwidth]{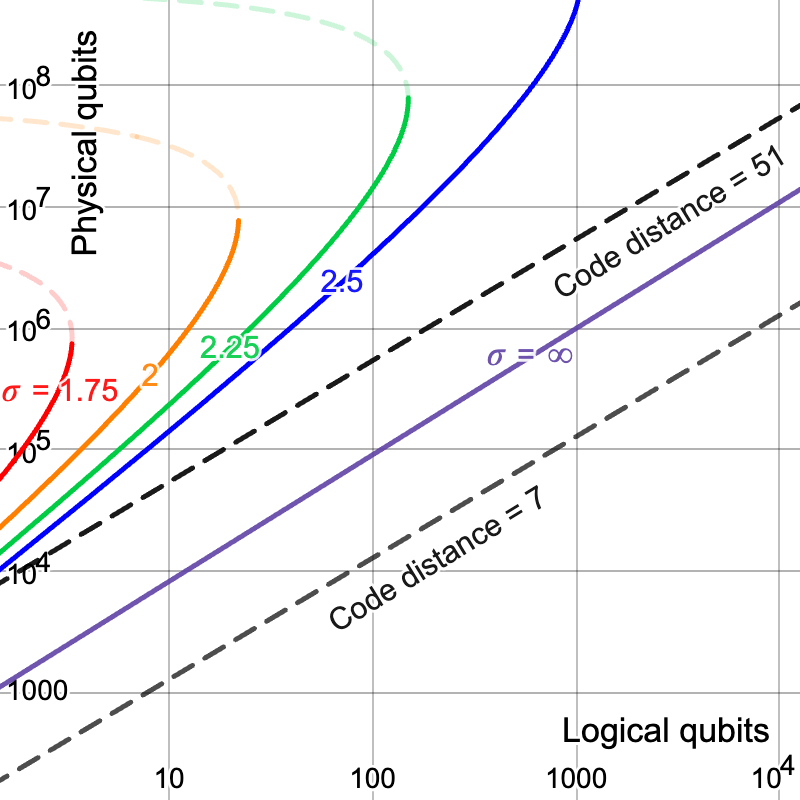}}
    \caption{This plot shows the QPE resource overhead under the logarithmic scalability model of Eq. \ref{eq:logscalability}. Similar to the power law model shown in Figure \ref{fig:scalability}, the logarithmic scalability model also predicts that, for each finite value of scalability parameter $\sigma$, there is a maximum problem instance size that can be accommodated by the architecture.
    However, the logarithmic scalability model is more optimistic in that, for the same base error rate $p_{\textup{phys}}(Q_{\textup{phys}}=1)=p_0$ and for the same (logarithmic) slope at $Q_{\textup{phys}}=1$, the maximum problem instance size is far larger for the logarithmic scalability model.
    Thus, in order to observe the limited problem instance size in the range of 10 to 10,000 logical qubits, we use the larger base error rate of $p_{0}=0.001$ compared to the $p_{0}=0.0001$ used in Figure \ref{fig:scalability}.
    An editable version of the plot can be accessed here: \href{https://www.desmos.com/calculator/cnh0vchq6l}{https://www.desmos.com/calculator/cnh0vchq6l}}
\label{fig:logscalability}
\end{figure}

\bibliographystyle{unsrt}
\bibliography{references}

\end{document}